\documentclass{article}
\usepackage[usenames,dvipsnames,svgnames,x11names]{xcolor}
\newcommand{\ysnoted}[1]{ {\textcolor{green} { ***TODO Later: #1 }}} 

\renewcommand{\ysnoted}[1]{} 

\usepackage[T1]{fontenc} 
\usepackage[normalem]{ulem} 

\usepackage{dsfont}

\newcommand{\hilite}[1]{{\textcolor{BlueViolet}{#1}}}

\usepackage[nocompress]{cite}
\usepackage{listings}

\lstdefinelanguage{XML}
{
basicstyle=\ttfamily\footnotesize,
  morestring=[b]",
  moredelim=[s][\bfseries\color{Maroon}]{<}{\ },
  moredelim=[s][\bfseries\color{Maroon}]{</}{>},
  moredelim=[l][\bfseries\color{Maroon}]{/>},
  moredelim=[l][\bfseries\color{Maroon}]{>},
  morecomment=[s]{<?}{?>},
  morecomment=[s]{<!--}{-->},
  commentstyle=\color{gray},
  stringstyle=\color{blue},
  identifierstyle=\color{red}
}
\usepackage{moreverb}

\usepackage[nounderscore]{syntax}

\usepackage[pdftex]{graphicx}
\graphicspath{{./figures/}}
\DeclareGraphicsExtensions{.pdf}
\usepackage[cmex10]{amsmath}
\usepackage{amssymb}
\usepackage{mathtools}
\usepackage{amsfonts}

\usepackage{bm}

\usepackage{subcaption}
\usepackage{makecell}

\usepackage{multirow} 
\usepackage{rotating} 
\usepackage{booktabs} 
\usepackage{colortbl} 
\usepackage{tablefootnote} 

\usepackage[pdftex,colorlinks=true,urlcolor=blue,citecolor=blue]{hyperref}

\usepackage{xspace}

\let\labelindent\relax
\usepackage{enumitem}

\usepackage{blindtext}

\begin{document}

\title{
Ocularone-Bench: Benchmarking DNN Models on GPUs to Assist the Visually Impaired}

\author{Suman Raj, Bhavani A Madhabhavi\thanks{\em were with Dream:Lab at the time of writing this paper.}, Kautuk Astu,\and Arnav A Rajesh\thanks{\em Equal Contribution}~\footnotemark[1],~ Pratham M\footnotemark[2]~\footnotemark[1]~ and Yogesh Simmhan\and
Department of Computational and Data Sciences,
\and Indian Institute of Science, Bangalore 560012 INDIA
\and Email: \{sumanraj, kautukastu, simmhan\}@iisc.ac.in}

\maketitle


\begin{abstract}
VIP navigation requires multiple DNN models for identification, posture analysis, and depth estimation to ensure safe mobility. Using a hazard vest as a unique identifier enhances visibility while selecting the right DNN model and computing device balances accuracy and real-time performance. We present Ocularone-Bench, which is a benchmark suite designed to address the lack of curated datasets for uniquely identifying individuals in crowded environments and the need for benchmarking DNN inference times on resource-constrained edge devices. The suite evaluates the accuracy-latency trade-offs of YOLO models retrained on this dataset and benchmarks inference times of situation awareness models across edge accelerators and high-end GPU workstations. Our study on NVIDIA Jetson devices and RTX 4090 workstation demonstrates significant improvements in detection accuracy, achieving up to $99.4\%$ precision, while also providing insights into real-time feasibility for mobile deployment. Beyond VIP navigation, Ocularone-Bench is applicable to senior citizens, children and worker safety monitoring, and other vision-based applications.
\end{abstract}

\section{Introduction}
Over 200 million people worldwide experience moderate to severe visual impairment, 
significantly impacting mobility and quality of life~\cite{whoStats}. Assistive technologies for \textit{Visually Impaired Persons (VIPs)} can enhance autonomy, confidence, and social inclusion. While voice-assisted smart canes~\cite{wewalk} and wearables provide sensor and video-based guidance, their limited range and Field of View (FoV) restrict hazard detection. 

Our prior work, \textit{Ocularone}~\cite{suman2023chi}, proposes a drone-based VIP assistance solution that can be coupled with handheld smartphones and edge accelerators to address these limitations. It leverages Computer Vision (CV) models for real-time visual analytics over videos from the front-facing cameras of ``buddy drones'' that follow the VIP, and offers alerts to enable their safe navigation in
%
complex environments.
This requires a suite of Deep Neural Network (DNN) models to accurately identify the VIP, analyze body posture to assess movement intent, and estimate depth for obstacle detection. A unique visual identifier, such as a \textit{hazard vest}, enhances reliability by ensuring precise recognition in diverse conditions. Given the real-time nature of such safety-critical applications, model accuracy is crucial to prevent misclassification. Also, selecting the appropriate DNN model and the compute device for inferencing is essential to balance accuracy and responsiveness. 

\begin{figure}
    \centering
    \subfloat[Diverse datasets]{
    \includegraphics[width=0.33\columnwidth]{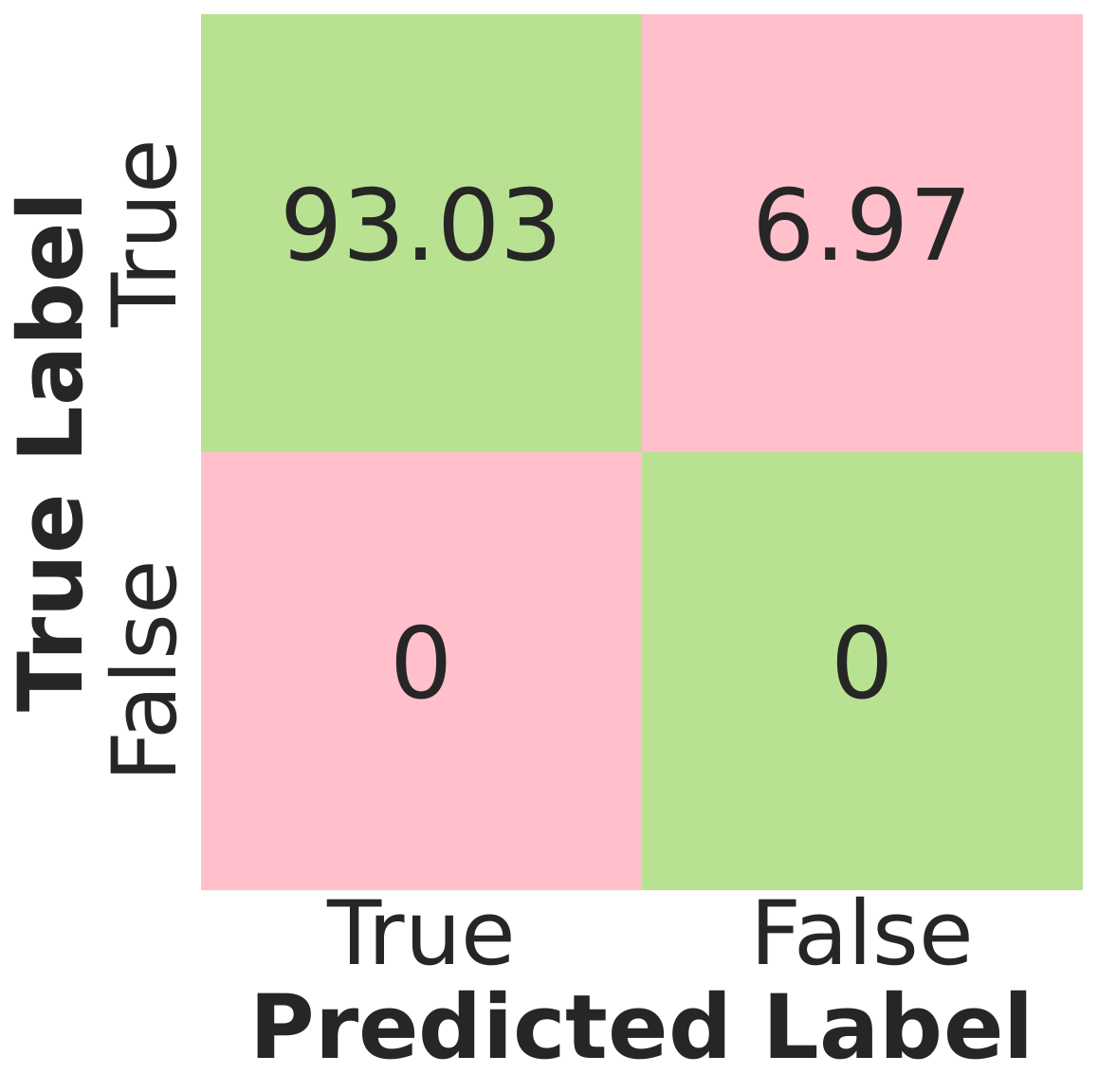}
    }\qquad
    \subfloat[Adversarial datasets]{
    \includegraphics[width=0.33\columnwidth]{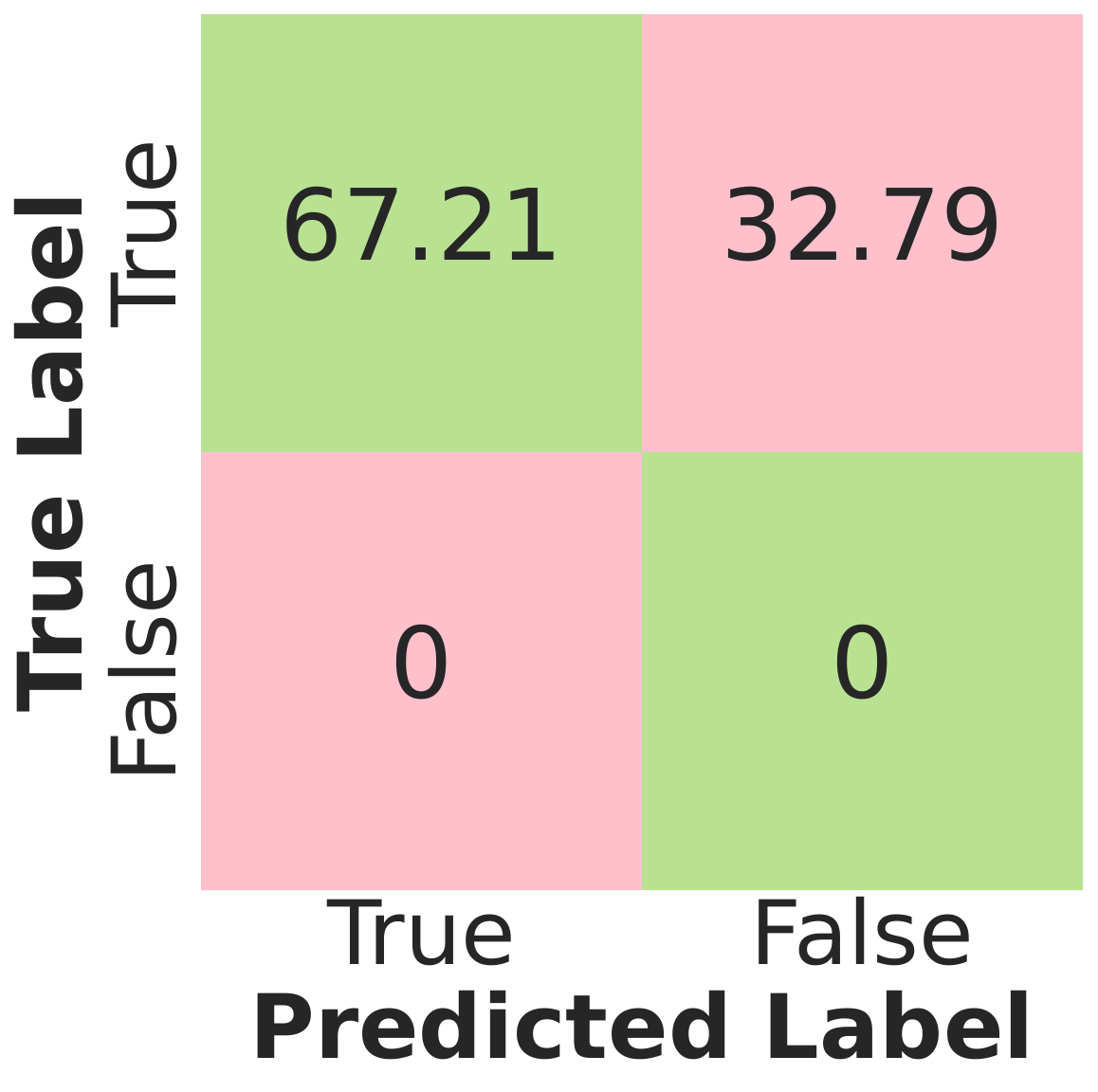}
    }\\ 
    \subfloat[Diverse datasets]{
    \includegraphics[width=0.33\columnwidth]{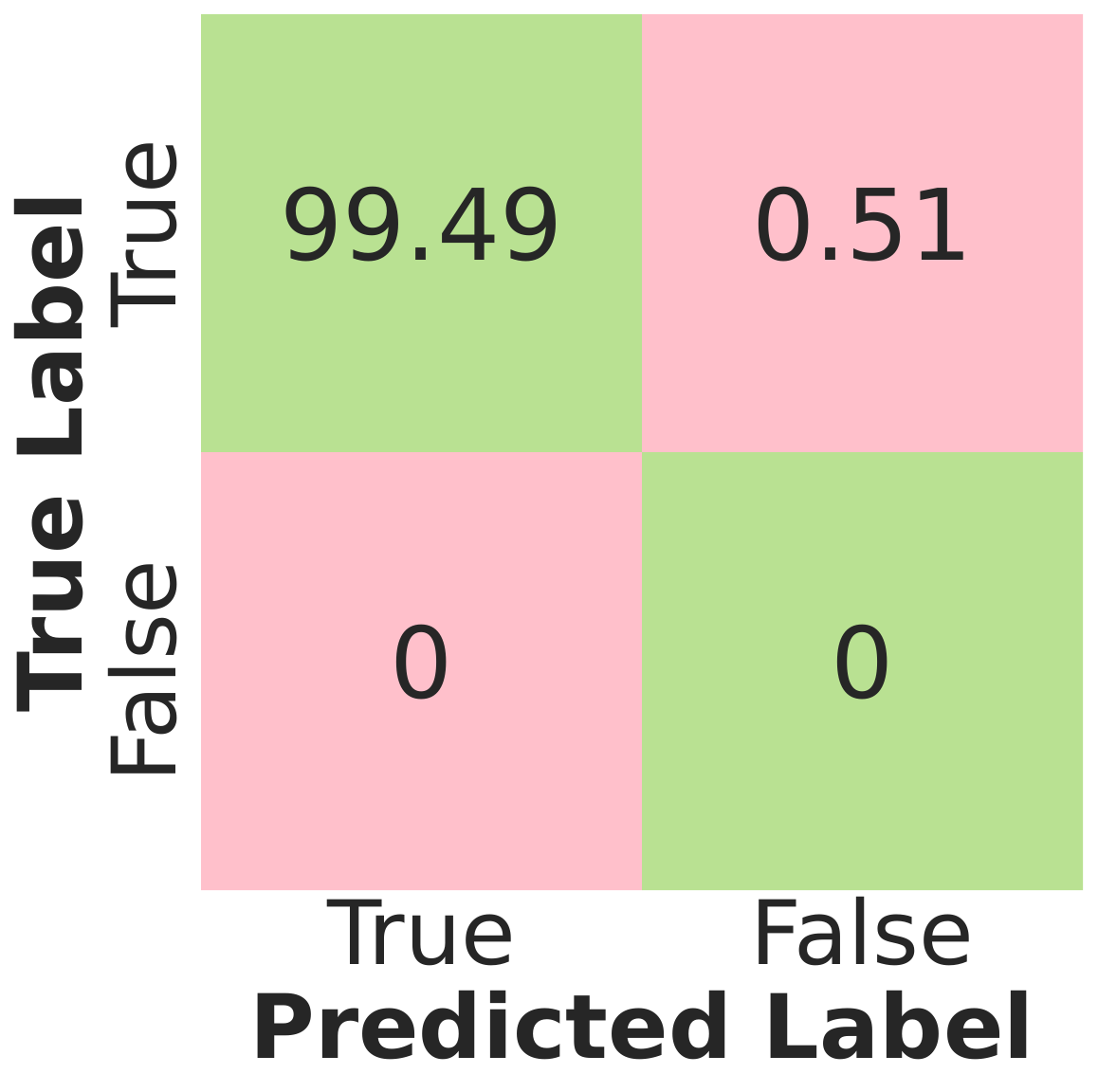}
    }\qquad
    \subfloat[Adversarial datasets]{
    \includegraphics[width=0.33\columnwidth]{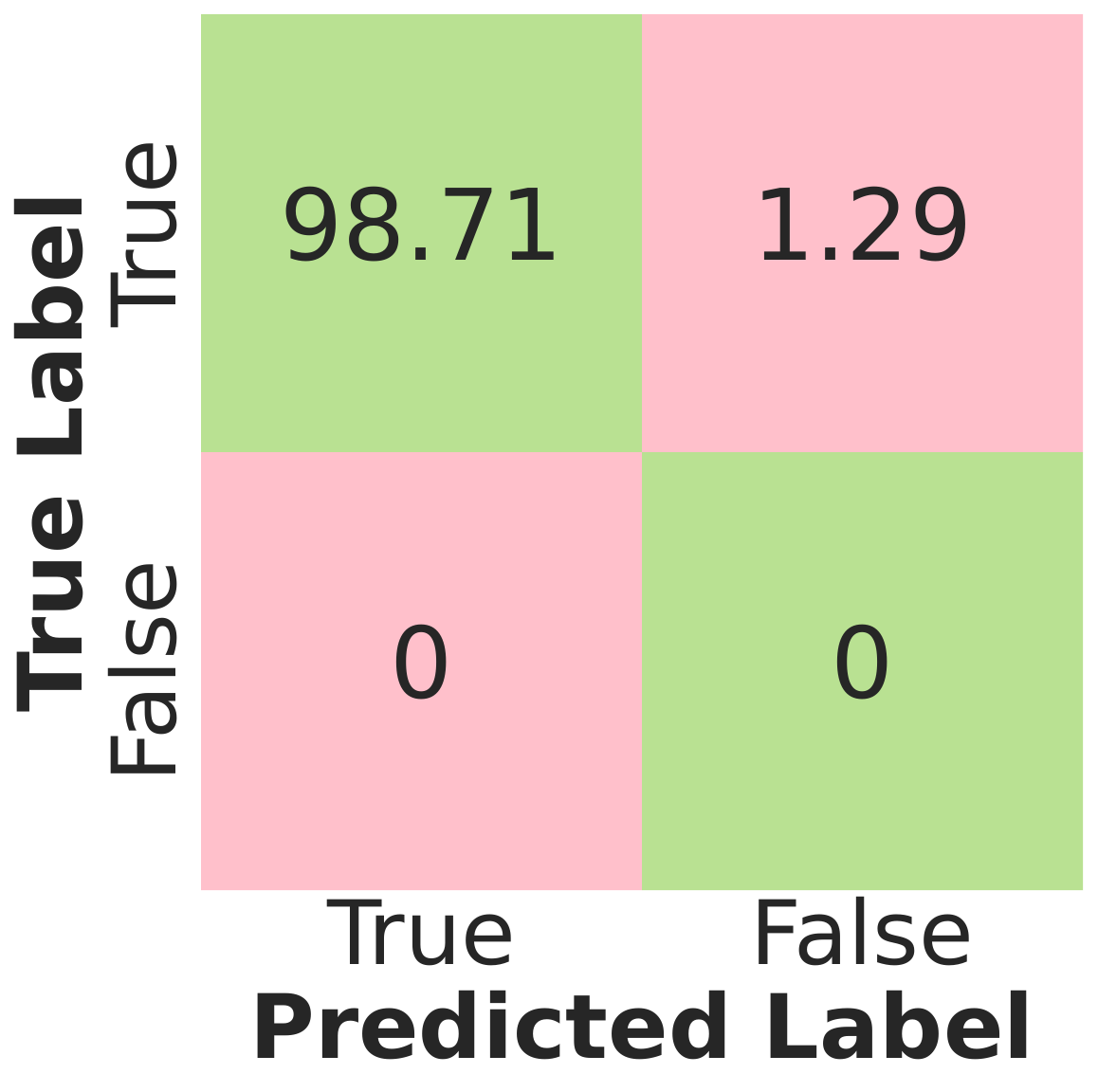}
    }
    \vspace{-0.05in}
    \caption{Accuracy of YOLOv11 (medium) trained using $1k$ random (top) and $3.8k$ curated (bottom) hazard-vest images}
    \label{fig:v11:comparison-with-old}
    \vspace{-0.2in}
\end{figure} 


\paragraph*{Challenges and Gaps} 
Identifying the VIP is one of the key tasks of VIP assistance systems. But DNN models for this task face challenges in uniquely identifying VIPs in crowded or dynamic environments. This is due to the lack of \textit{curated datasets} to train these models upon in diverse conditions. A review of top Hazard Vest (HV) image datasets and DNN models~\cite{roboflow_vest_dataset} 
reveals these gaps.
E.g., the SH-17~\cite{ahmad2024sh17} benchmark reports a peak precision of $81\%$ for a generic YOLOv9-e model while a YOLOv8-s model trained on 795 HV images improves this to $85.7\%$ precision~\cite{hazard-vest_dataset}. In our study (Fig.~\ref{fig:v11:comparison-with-old}) we achieve $93\%$ precision on a YOLOv11-m retrained on a dataset of $1k$ HV images
whereas retraining it on a curated set of $3.8k$ images improves the precision to $99.5\%$.
This highlights the impact of dataset size and quality on model performance. 

Further, existing DNN benchmarks report inference times of common models on some edge devices~\cite{10.1007/978-981-96-0805-8_11}, but fail to offer a diverse set of performance numbers of relevant DNN models on target edge accelerators used for VIP assistance.

\paragraph*{Contributions} 
In this paper, we address these limitations and introduce \textit{Ocularone-Bench}\footnote{\url{https://github.com/dream-lab/ocularone-dataset}}, a benchmark suite that offers a curated dataset for hazard vest detection, achieving up to $99.4\%$ precision. Additionally, we benchmark the inference times of these models on multiple edge accelerators and GPU workstation, along with performance of other situational awareness DNN models. While developed for VIP navigation, this dataset, and retrained models are versatile and applicable to broader domains, such as safety monitoring of senior citizens, children and worker.

We make the following key contributions: 
\begin{enumerate}[leftmargin=*]
    \item We curate an annotated dataset of $30k$ images of a person wearing hazard vest in diverse outdoor conditions (\S~\ref{sec:dataset}).
    \item  We retrain various sizes of state-of-the-art object detection models, YOLOv8 and YOLOv11 (\S~\ref{sec:vip-application-models}). We offer an detailed analysis of accuracy vs. latency tradeoffs on accelerated edge devices and a GPU workstation (\S~\ref{sec:evaluation}). 
    \item Lastly, we report inference times of diverse situation awareness DNN models used for VIP assistance on these devices. 
\end{enumerate}
We also offer our conclusions and outline potential directions for future research in \S~\ref{sec:conclusions}.

\section{Ocularone Dataset Description} \label{sec:dataset}

\begin{table}[!t]
    \centering
    \footnotesize
    \setlength{\tabcolsep}{2pt}
    \caption{Dataset Summary}
    \begin{tabular}{l|l|r}
    \hline
    \textbf{Category} & \textbf{Sub-Category} & \textbf{\makecell{\# of \\annotated \\images}}\\
       \hline
       \hline 
         & a. No pedestrians & 2294 \\
        \cline{2-3}
        1. Footpath & b. Pedestrians in FoV & 1371 \\
       \cline{2-3}
         & c. Usual surroundings & 2115\\ 
        \hline
         & a. Bicycles in FoV & 901 \\
        \cline{2-3}
        2. Path & b. Pedestrians in FoV & 1658\\
        \cline{2-3}
        & c. Pedestrians \& Cycles in FoV & 1057\\
        \hline
        & a. Pedestrians in FoV & 1326\\
        \cline{2-3}
        3. Side of road  & b. Usual Surroundings & 1887 \\
        \cline{2-3}
        & c. No pedestrians in FoV & 2022 \\
        \cline{2-3}
         & d. Parked cars in FoV & 2527 \\
        \hline
        4. Mixed scenarios &  & 9169 \\
        \hline
        5. Adversarial scenarios & Low light, blur, cropped image, etc. & 4384\\
        \hline
        \hline
        \textbf{Total} & & \textbf{30711} \\ 
        \hline 
    \end{tabular}
    \label{tab:diverse-dataset}
\end{table}

\begin{figure}[!t]
\centering
    \subfloat[Category 4]{
    \includegraphics[width=0.28\columnwidth]{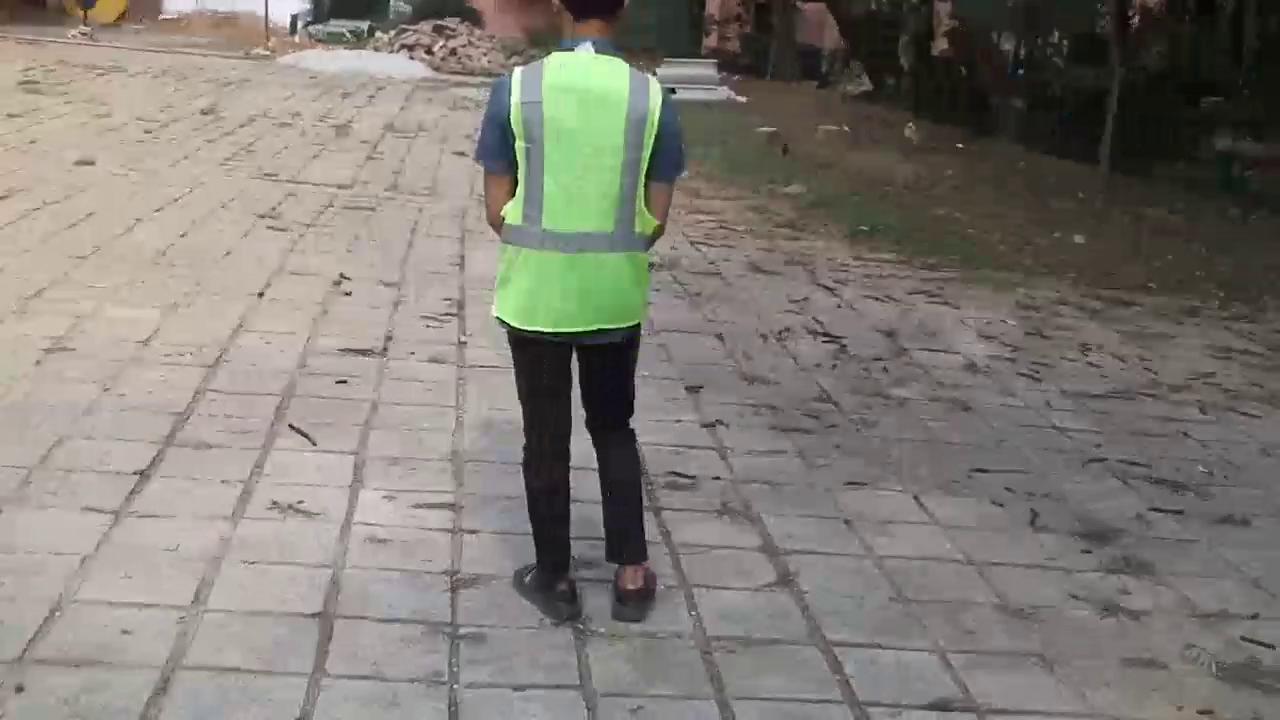}
   \label{fig:sample-1}
  }\hfill
  \subfloat[Category 1.(a)]{
    \includegraphics[width=0.28\columnwidth]{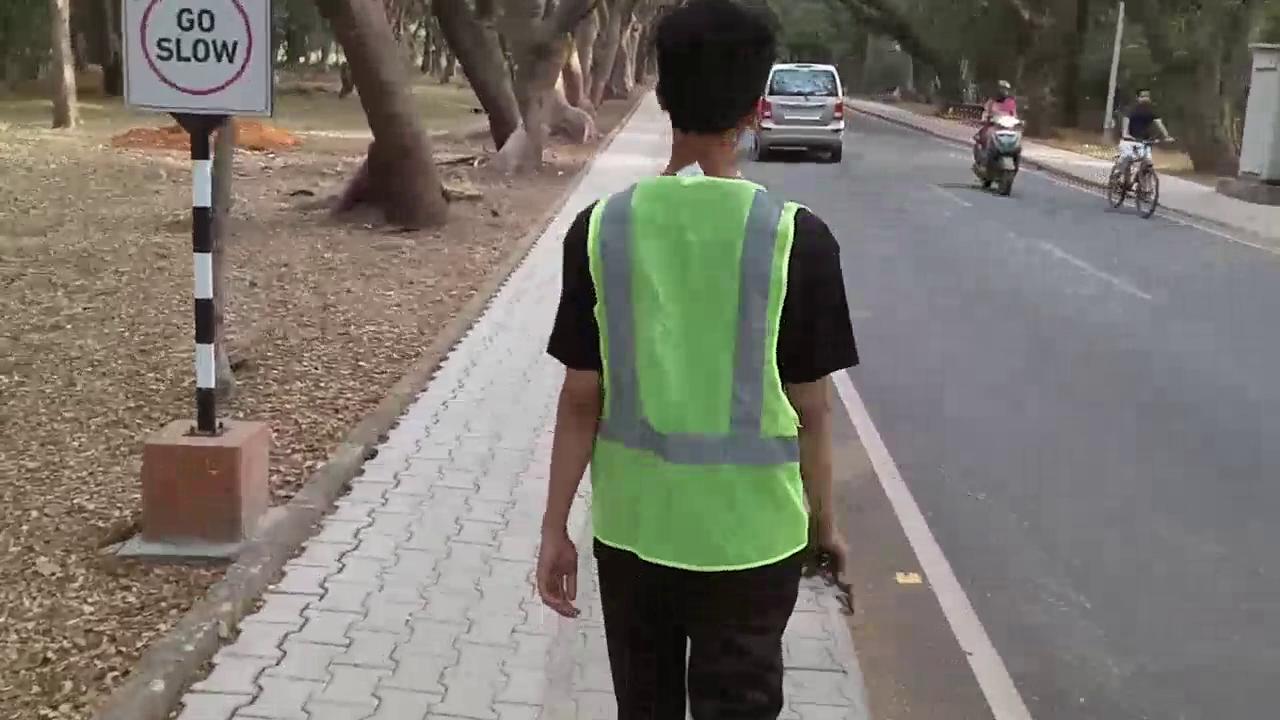}
   \label{fig:sample-2}
  }\hfill
  \subfloat[Category 1.(c)]{
   \includegraphics[width=0.28\columnwidth]{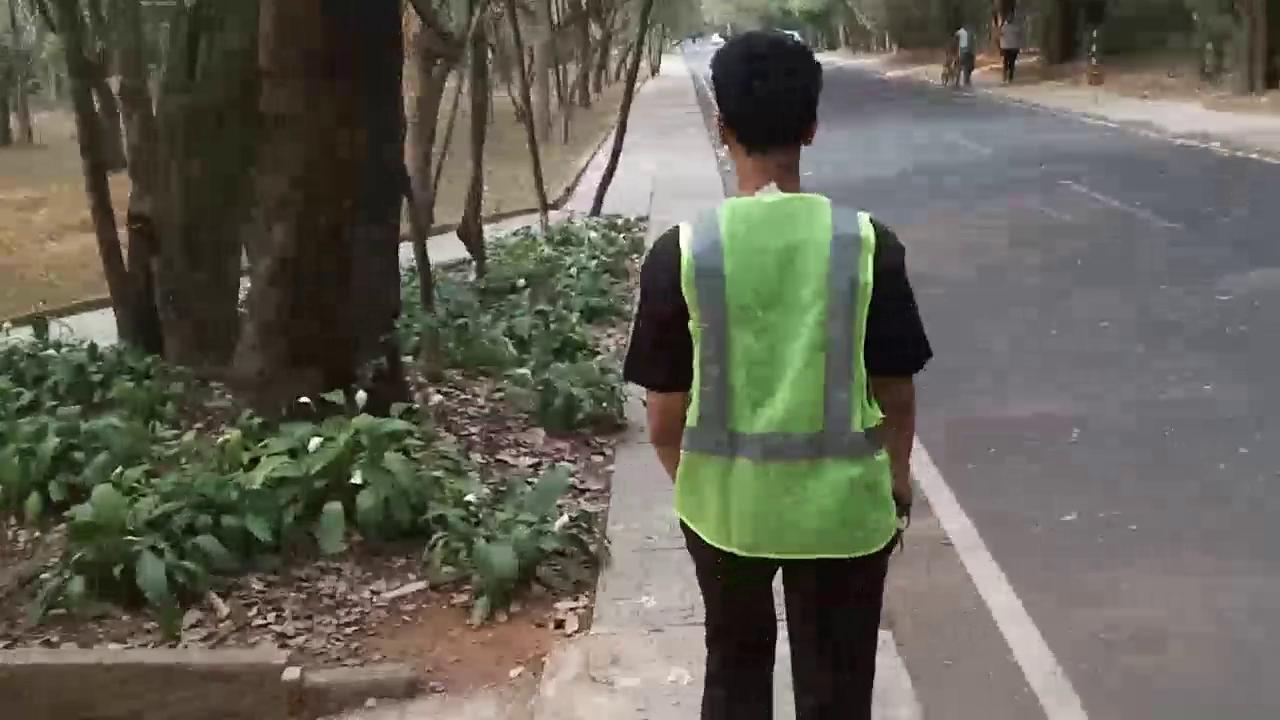}
    \label{fig:sample-3}
  }\\
   \subfloat[Category 1.(b)]{
    \includegraphics[width=0.28\columnwidth]{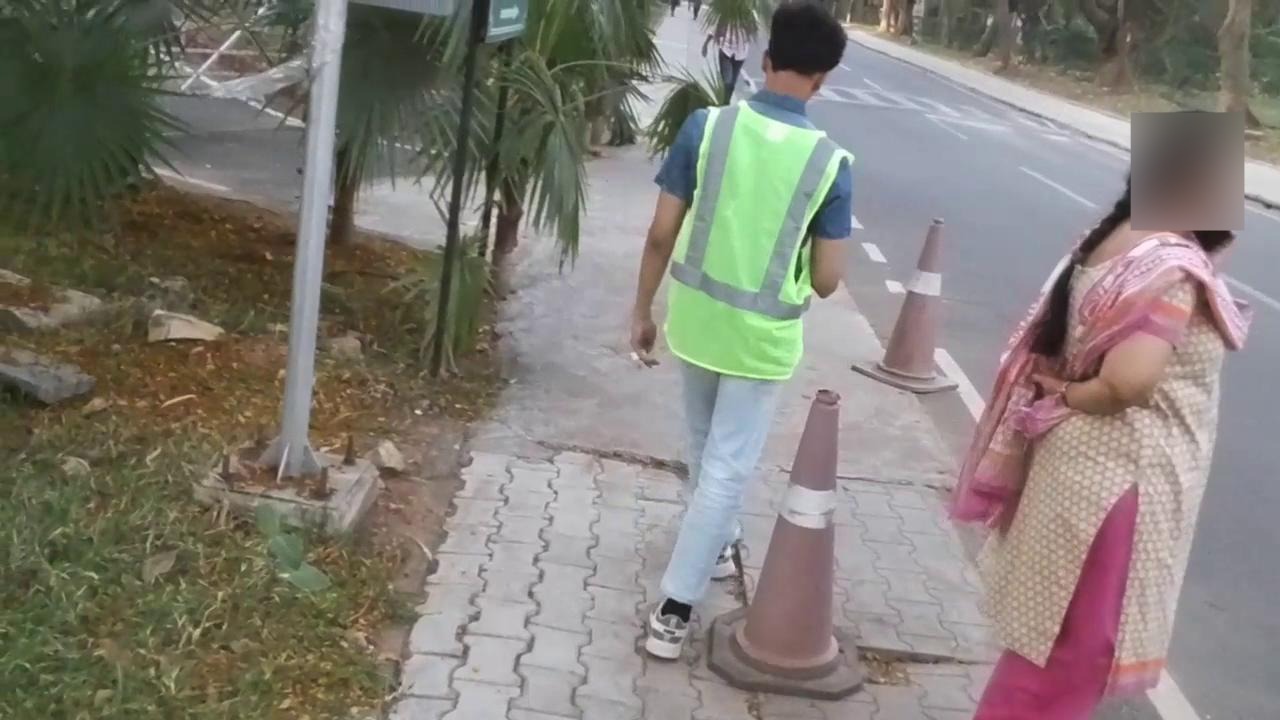}
   \label{fig:sample-4}
  }\hfill
  \subfloat[Category 2.(c)]{
    \includegraphics[width=0.28\columnwidth]{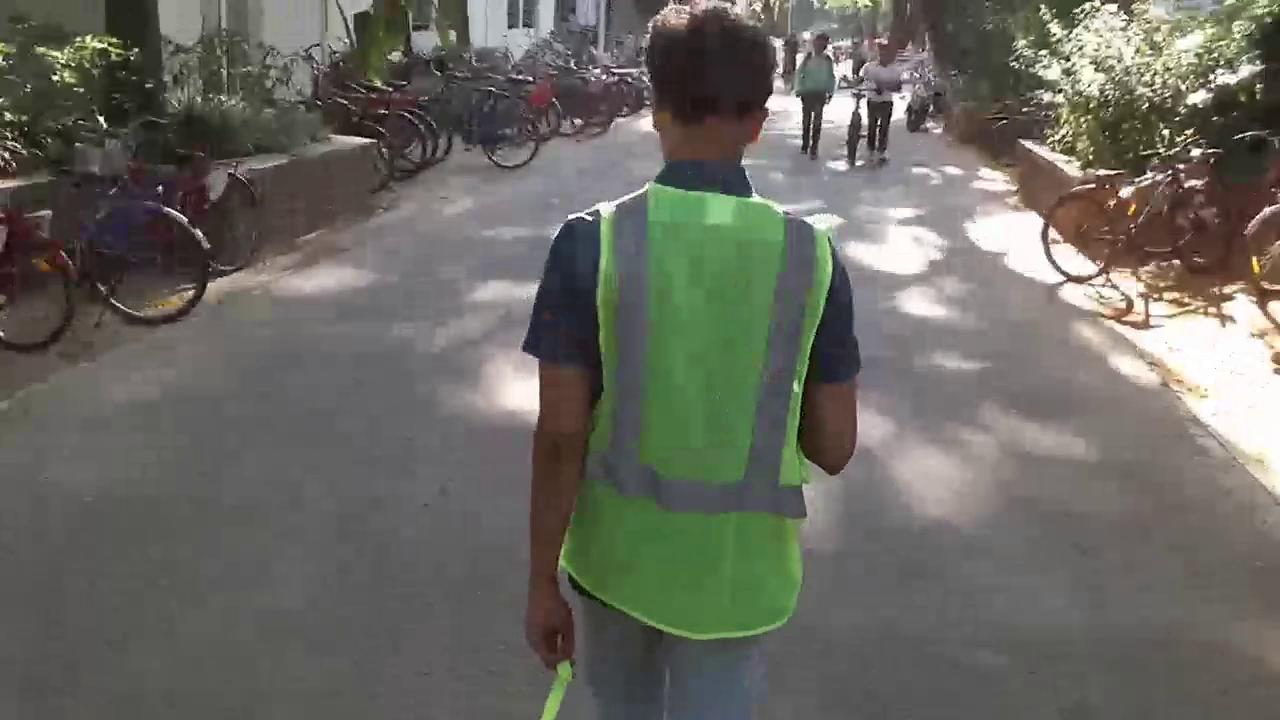}
   \label{fig:sample-5}
  }\hfill
  \subfloat[Category 3.(a)]{
   \includegraphics[width=0.28\columnwidth]{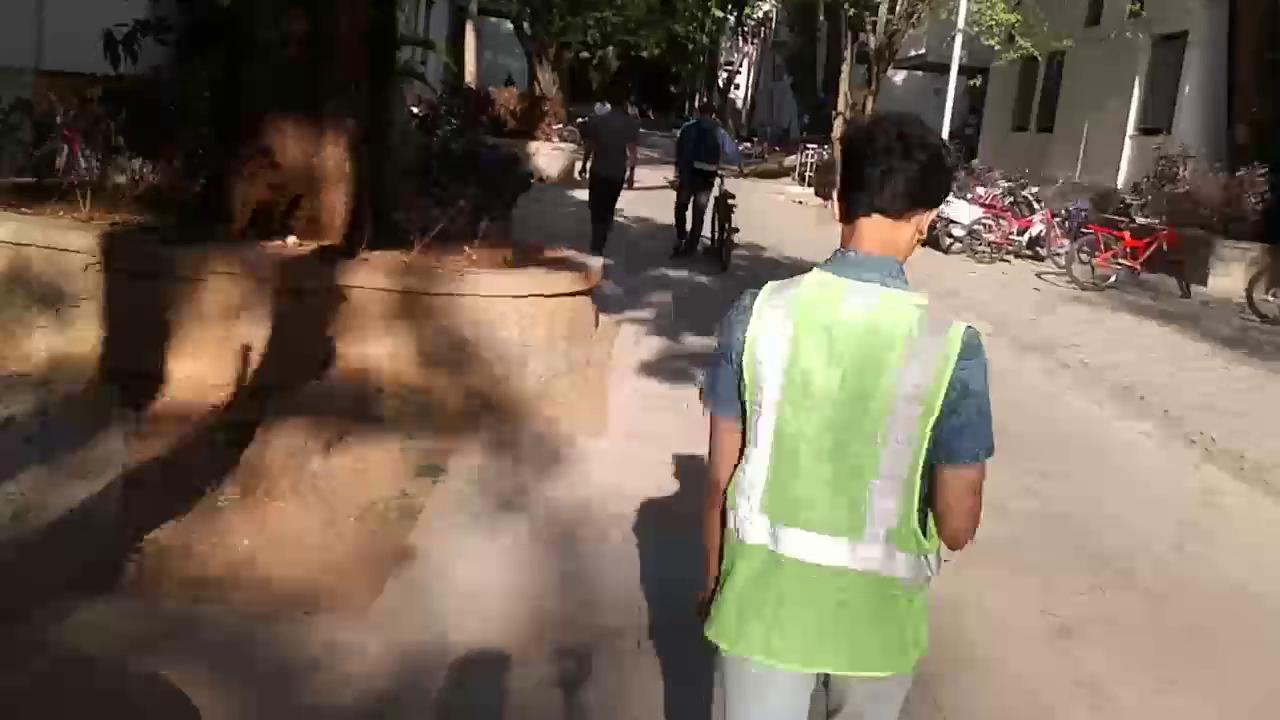}
    \label{fig:sample-6}
  }
\caption{Sample images from the dataset}
\label{fig:sample-data}
\end{figure}

We collect a total of $43$ videos of duration between $1-2$~minutes at different locations in our university campus. The videos were recorded using a DJI Tello nano quad-copter which has an onboard $720$p HD monocular camera that generates feeds at $30$ frames per second (FPS). The drone was handheld at different heights and distances while following the proxy VIP — who wore a hazard vest — around our university campus. To extract frames from these videos, we used the moviepy library\footnote{https://pypi.org/project/moviepy/} in Python, which supports a wide range of media processing tasks, including video editing and frame extraction. Specifically, the \textit{editor} module of moviepy was utilized to extract frames at $10$~FPS. This generated a dataset of 30,711 images capturing a proxy VIP walking through various real-world scenarios. 

Table \ref{tab:diverse-dataset} presents a summary of the dataset which is categorized based on different scenarios in which the VIP walks, including footpaths, paths, and the side of the road, with sub-categories specifying the presence of pedestrians, bicycles, parked cars, and usual surroundings. Additionally, mixed scenarios, which include a combination of these conditions, contribute $\approx9k$ images. These reflect real-world navigation scenarios for VIPs in outdoor environments, where accurate hazard detection is critical and presents varying levels of obstacles, textures, and lighting conditions, making them essential for training robust models to aid practical deployment. The dataset also includes 4,384 images captured under adversarial conditions like low light, blur, cropping, and tilted orientations to enhance robustness. These diverse visuals support not only our application but also future research in pedestrian detection, path navigation, and drone-based scene understanding. Some samples of this datasets are shown in Fig.~\ref{fig:sample-data}.
Finally, these datasets are annotated in Roboflow by drawing a bounding box around the region of interest, the "neon hazard vest", using the \href{https://www.makesense.ai/}{``makesense.ai''} tool. The Roboflow annotation file includes the class label of the image, along with the top-left and bottom-right coordinates of the bounding box.

\section{VIP Application Specific DNN  Models}\label{sec:vip-application-models}\label{sec:dnn-models}

\begin{table}[t]
\centering
\caption{Specifications of DNN Models considered for Ocularone-Bench}
\footnotesize
\setlength{\tabcolsep}{2pt}
\begin{tabular}{l|c|c|r|r}
\hline
\textbf{Category} & \textbf{Architecture} & \textbf{Model} & \bf{\makecell{\# of \\parameters\\(in millions)}} & \bf{\makecell{Model Size \\(in MB)}} \\
\hline\hline
 & & v8-n & 3.2 & 5.95 \\ 
Vest Detection & YOLO & v8-m & 25.9 & 49.61 \\
 & & v8-x & 68.2 & 130.38 \\
\hline
 & & v11-n & 2.6 & 5.22 \\ 
Vest Detection & YOLO & v11-m & 20.1 & 38.64 \\
 & & v11-x & 56.9 & 109.09 \\
\hline
Pose Detection & ResNet-18 & trt\_pose & 12.8 & 25 \\ 
\hline
Depth Estimation & ResNet-18 & Monodepth2 & 14.84 & 98.7 \\
\hline
\end{tabular}
\label{tab:dnn-model-specs}
\end{table} 

For our VIP application, we incorporate multiple DNN models used in~\cite{raj2024adaptiveheuristicsschedulingdnn}. We select YOLO~\cite{ultralytics2025} models, specifically YOLOv8 and YOLOv11, which we retrain to detect hazard vests. Instead of using all available YOLO model sizes, we strategically choose three specific size variants — Nano (n), Medium (m), and X-Large (x) — to effectively cover the spectrum of trade-offs between lightweight, real-time inference on edge devices (n), balanced performance (m), and high-accuracy detection with greater computational demands (x). Compared to other models like Faster R-CNN, which uses a two-stage detector, YOLO's single-shot detection framework enables faster inference. These make it well-suited for edge deployment where quick and reliable VIP identification is essential for real-time mobility assistance. Additionally, we have an out-of-the-box body pose estimation model~\cite{nvidia_trt_pose}, which helps evaluate the VIP’s posture and movement. This is integrated with an SVM classifier to detect \textit{fall} scenarios. 

Beyond object and pose detection, we use Monodepth2~\cite{monodepth2github} for depth estimation, providing spatial awareness crucial for obstacle avoidance and path planning. Together, these models enhance VIP assistance by integrating object detection, pose estimation, and depth perception for safer navigation. Table~\ref{tab:dnn-model-specs} summarizes the models used in our benchmarks.

\subsection{Retraining of YOLO models}

We randomly sample $\approx10\%$ images from each of the scene category and use a total of $3,866$ images from $12$ different categories as training data, while the remaining images are set aside for testing the re-trained model. The training data is further split into an $80:20$ ratio, with $20\%$ serving as the validation dataset. The final training and validation datasets are uploaded to Roboflow, a platform for building and deploying computer vision models, to generate a YAML file required for training the YOLOv8 and YOLOv11 model. We have used the default parameters provided by Ultralytics, with a learning rate of $0.01$ and an IoU (Intersection over Union) threshold of $0.7$. Both models are trained on a fixed image size of $640\times640$ in batch of $16$ for a total of $100$ epochs.
The labelled dataset, trained models, and inference scripts are publicly available on our GitHub repository.

\section{Evaluation}\label{sec:evaluation}

\begin{table}[t]
\centering
\caption{Specifications of NVIDIA Jetson edge computing devices used in evaluations}
\footnotesize
\setlength{\tabcolsep}{2pt}
\begin{tabular}{l|r|r|r}
\hline
\textbf{Feature} & \bf {Orin AGX} & \bf {Xavier NX } &\bf{Orin Nano} \\
\hline\hline
GPU Architecture & Ampere & Volta & Ampere\\
\noalign{\global\arrayrulewidth=0.1pt}\arrayrulecolor{lightgray} \hline
\noalign{\global\arrayrulewidth=0.4pt}\arrayrulecolor{black}
$\#$ CUDA/Tensor Cores & 2048/64 & 384/48 & 1024/32\\
\noalign{\global\arrayrulewidth=0.1pt}\arrayrulecolor{lightgray} \hline
\noalign{\global\arrayrulewidth=0.4pt}\arrayrulecolor{black}
RAM (GB) & $32$ & $8$ & $8$ \\ \hline
Jetpack Version & 6.1 & 5.0.2 & 5.1.1 \\
\noalign{\global\arrayrulewidth=0.1pt}\arrayrulecolor{lightgray} \hline
\noalign{\global\arrayrulewidth=0.4pt}\arrayrulecolor{black}
CUDA Version & 12.6 & 11.4 & 11.4 \\ \hline 
\noalign{\global\arrayrulewidth=0.1pt}\arrayrulecolor{lightgray} \hline
\noalign{\global\arrayrulewidth=0.4pt}\arrayrulecolor{black}
Peak Power (W) & 60 & 15 & 15\\ \hline
Form factor (mm) & $110 \times 110 \times 72$ & $103 \times 90 \times 35$ & $100 \times 79 \times 21 $\\ 
\noalign{\global\arrayrulewidth=0.1pt}\arrayrulecolor{lightgray} \hline
\noalign{\global\arrayrulewidth=0.4pt}\arrayrulecolor{black}
Weight (g) & 872.5 & 174 & 176\\ \hline 
Price (USD) & $\$2370$  & $\$460$ & $\$630$\\ 
\hline
\end{tabular}
\label{tab:computespecs}
\end{table}

\subsection{Setup} 

We implement our benchmark scripts in \textit{Python}. All inferencing experiments were run on three NVIDIA Jetson edge devices and one high-end GPU workstation. The technical specifications of the edge devices have been shared in Table~\ref{tab:computespecs} and we use NVIDIA RTX $4090$ as the GPU workstation, which has $16,384$~CUDA core NVIDIA GPU based on Ampere architecture with $512$ tensor cores, a AMD Ryzen 9 7900X 12-Core Processor CPU and a $24$GB GPU RAM. The training was run independently on an NVIDIA A5000 GPU. We use \textit{PyTorch $2.0.0$} for invoking the various DNN models for inferencing over the images. 

\subsection{Results}

\begin{figure}[!t]
\centering
    \subfloat[Nano]{
    \includegraphics[width=0.28\columnwidth]{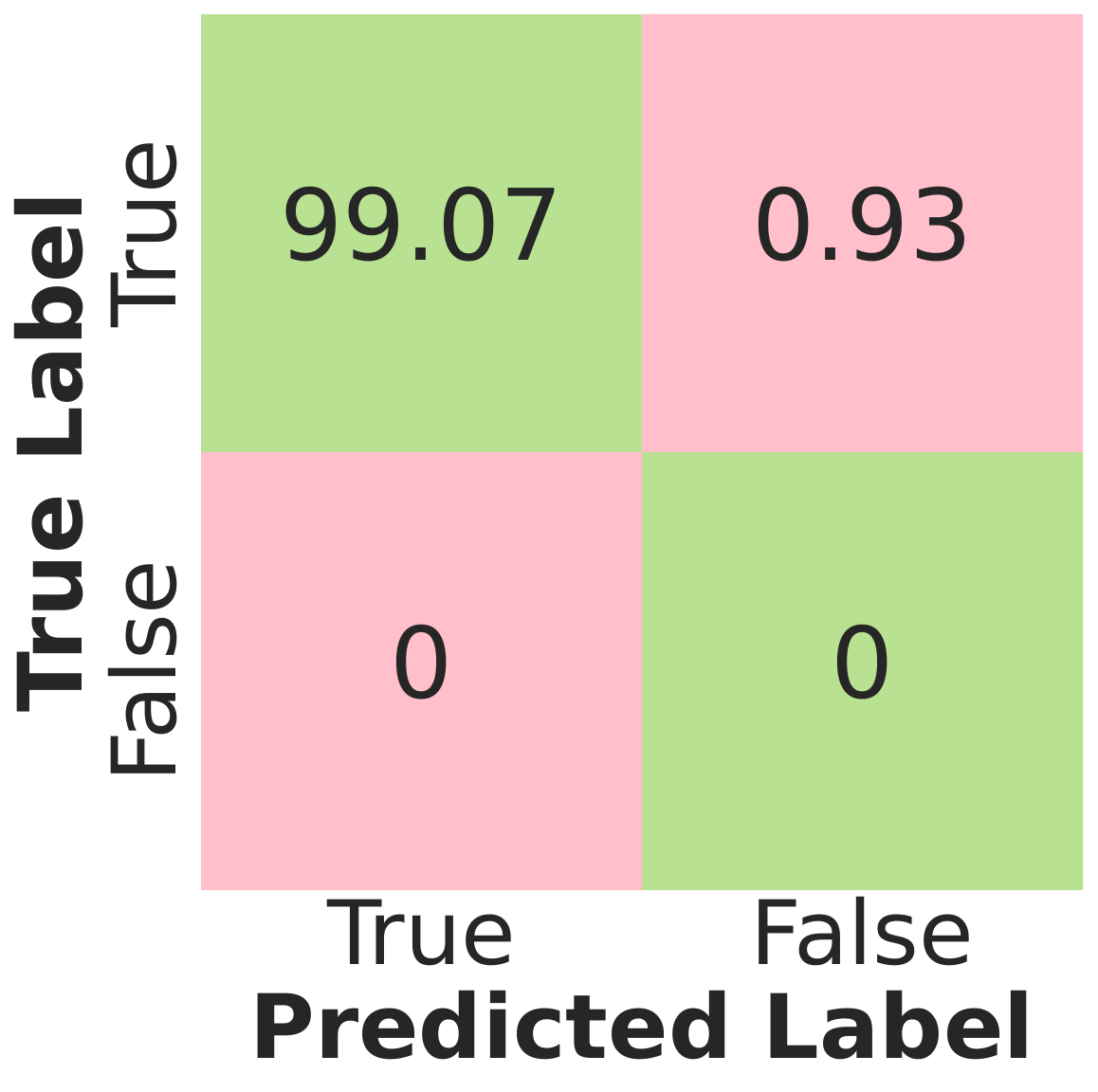}
   \label{fig:yolov8-1-diverse}
  }~~
  \subfloat[Medium]{
    \includegraphics[width=0.28\columnwidth]{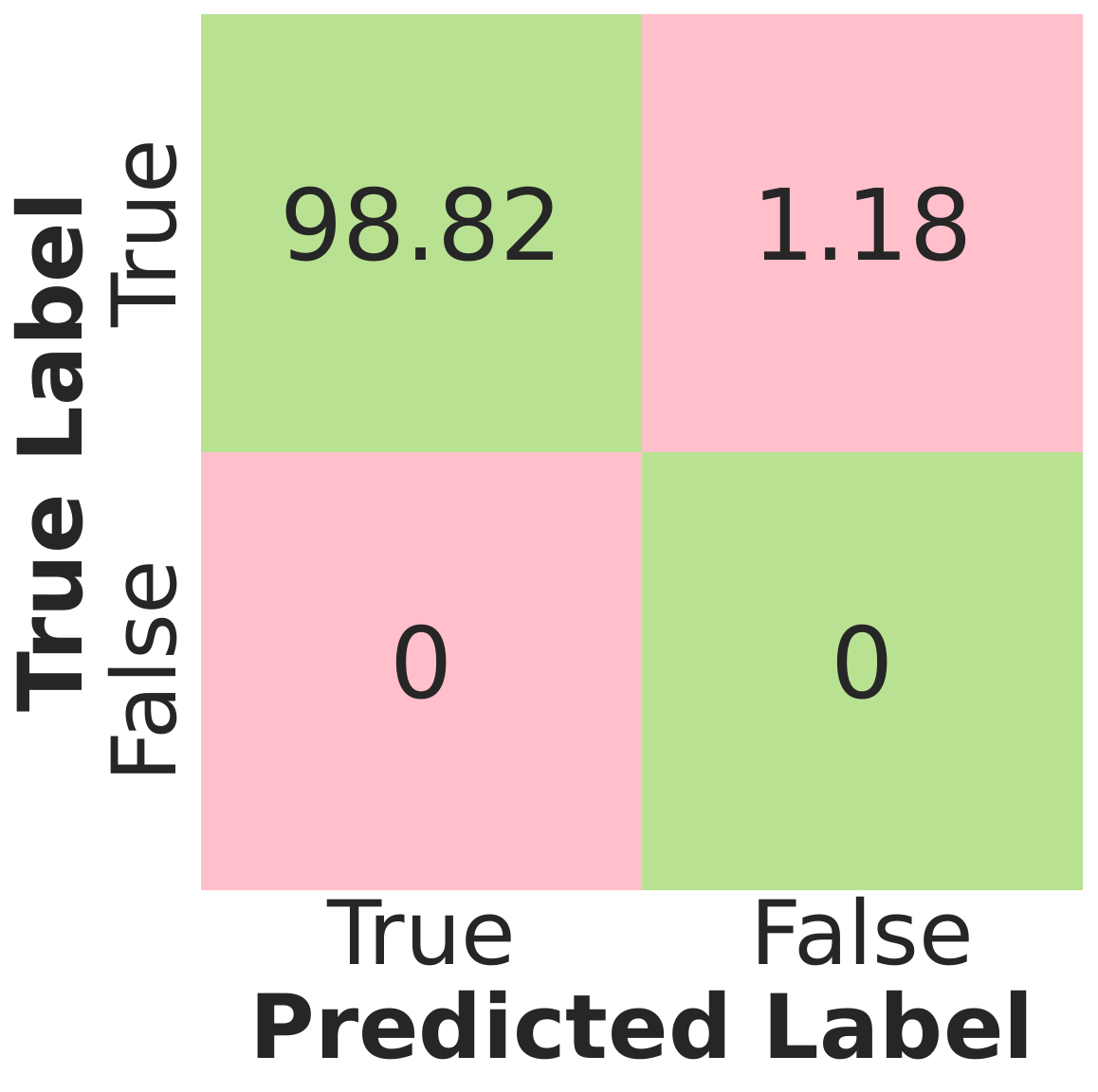}
   \label{fig:yolov8-2-diverse}
  }~~
  \subfloat[X-Large]{
   \includegraphics[width=0.28\columnwidth]{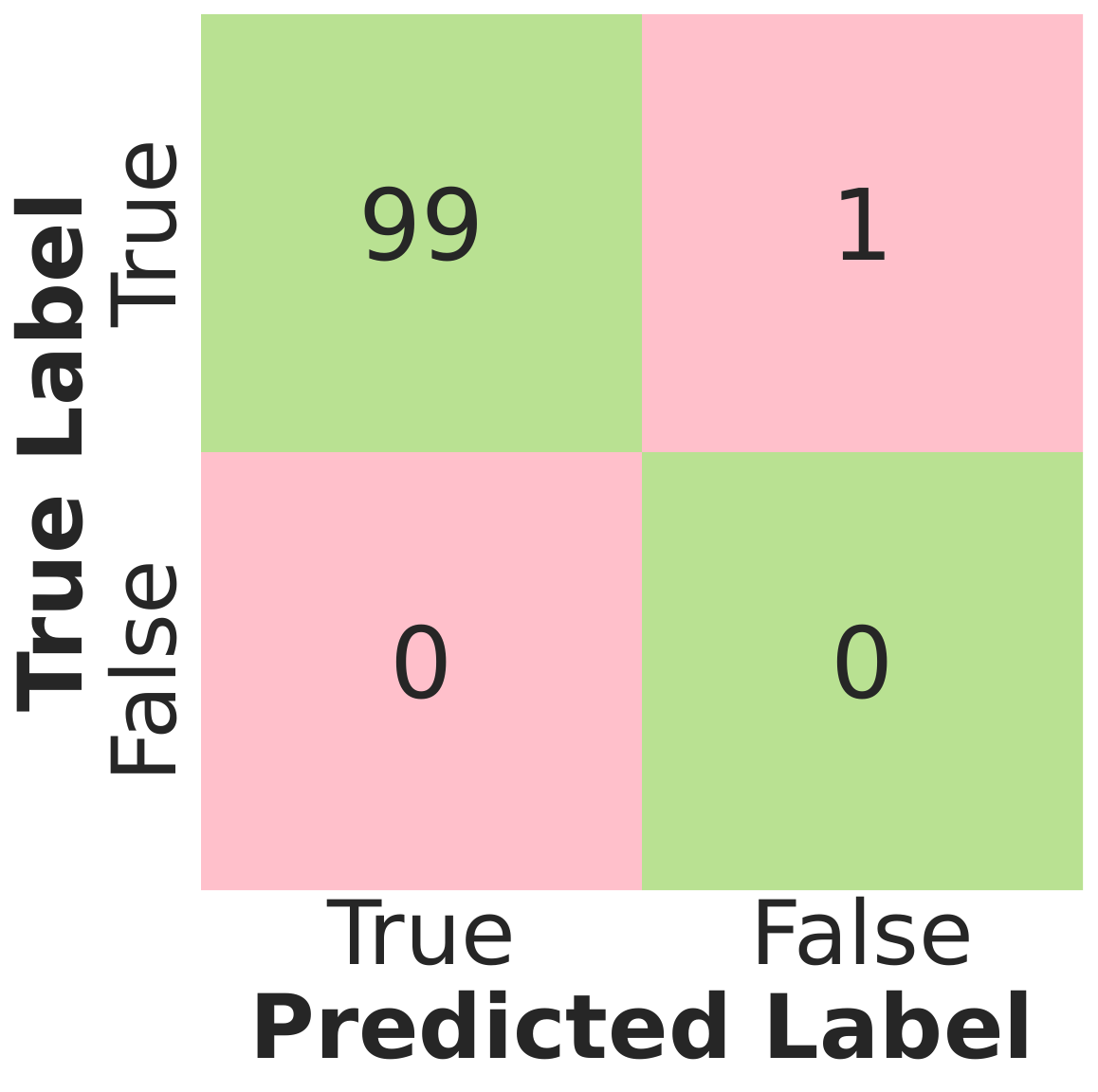}
    \label{fig:yolov8-3-diverse}
  }\\
   \subfloat[Nano]{
    \includegraphics[width=0.28\columnwidth]{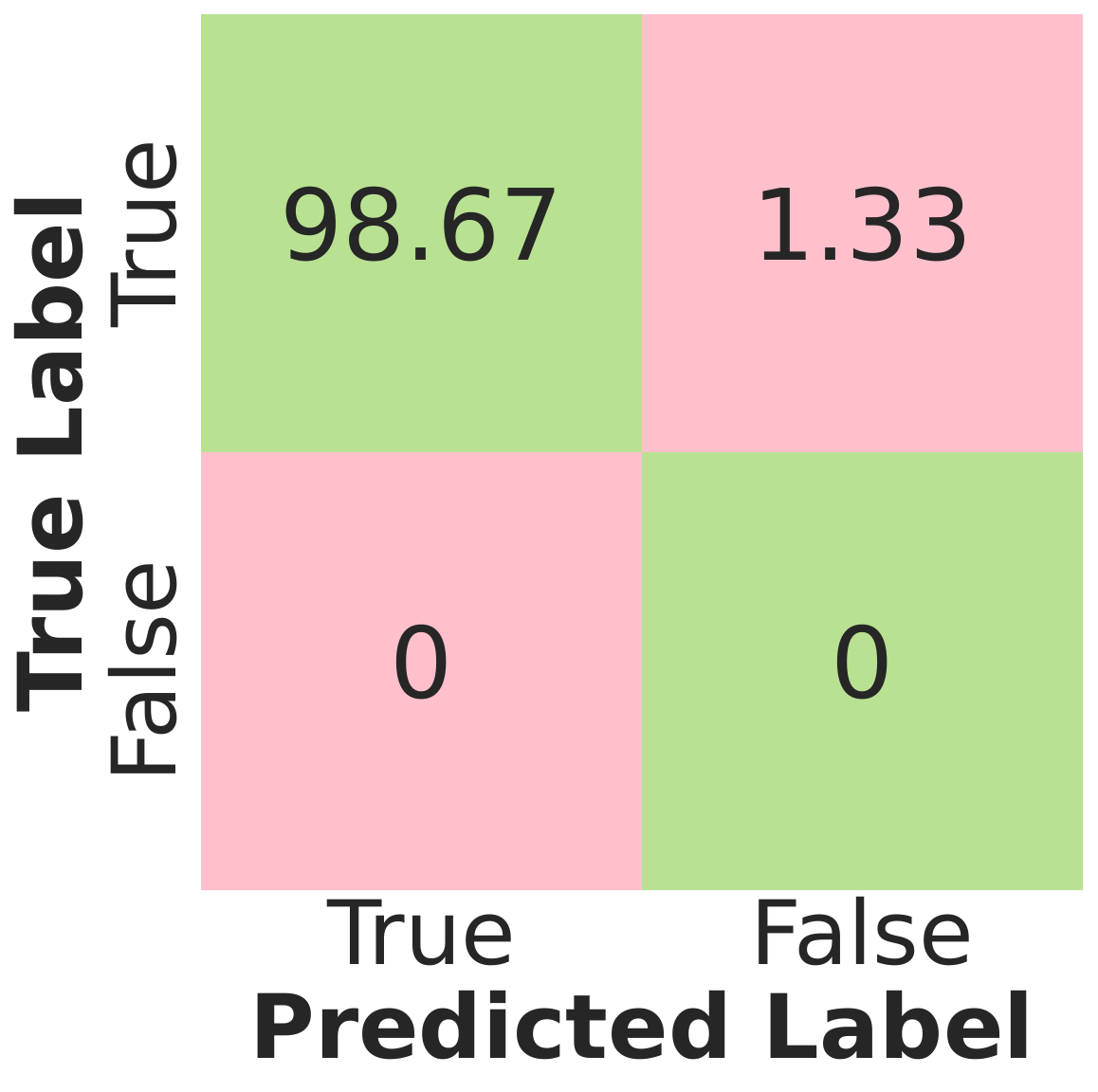}
   \label{fig:yolov11-1-diverse}
  }~~
  \subfloat[Medium]{
    \includegraphics[width=0.28\columnwidth]{Scratch-Figs/diverse/custom_yolov11_m_100_epochs.pdf}
   \label{fig:yolov11-2-diverse}
  }~~
  \subfloat[X-Large]{
   \includegraphics[width=0.28\columnwidth]{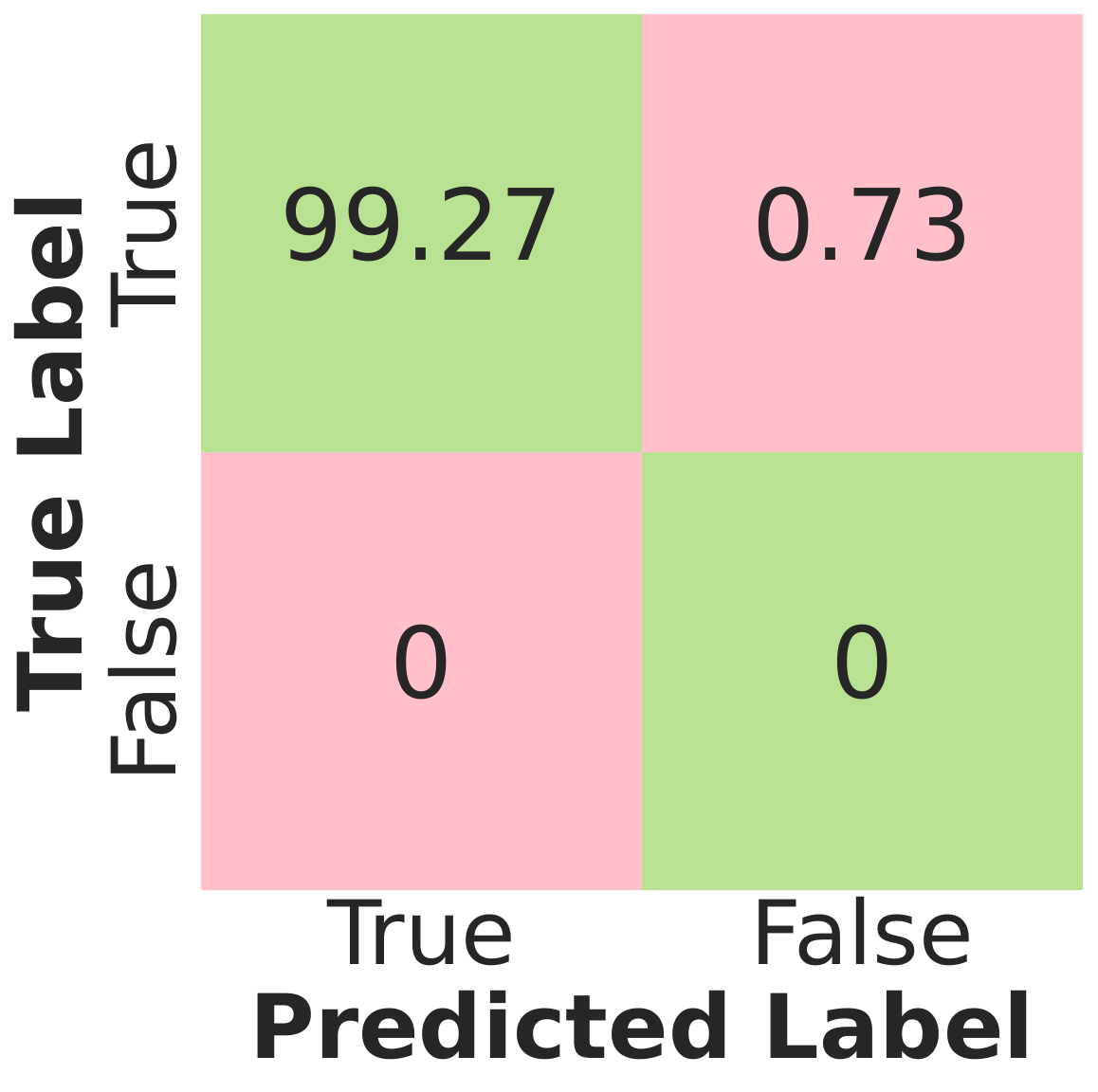}
    \label{fig:yolov11-3-diverse}
  }
\caption{Accuracy (in \%) of VIP detection using different sizes of Re-trained (RT) YOLOv8 (top) and YOLOv11 (bottom) on diverse datasets}
\label{fig:diverse-accuracy}
\end{figure}


\begin{figure}[!t]
\centering
    \subfloat[Nano]{
    \includegraphics[width=0.28\columnwidth]{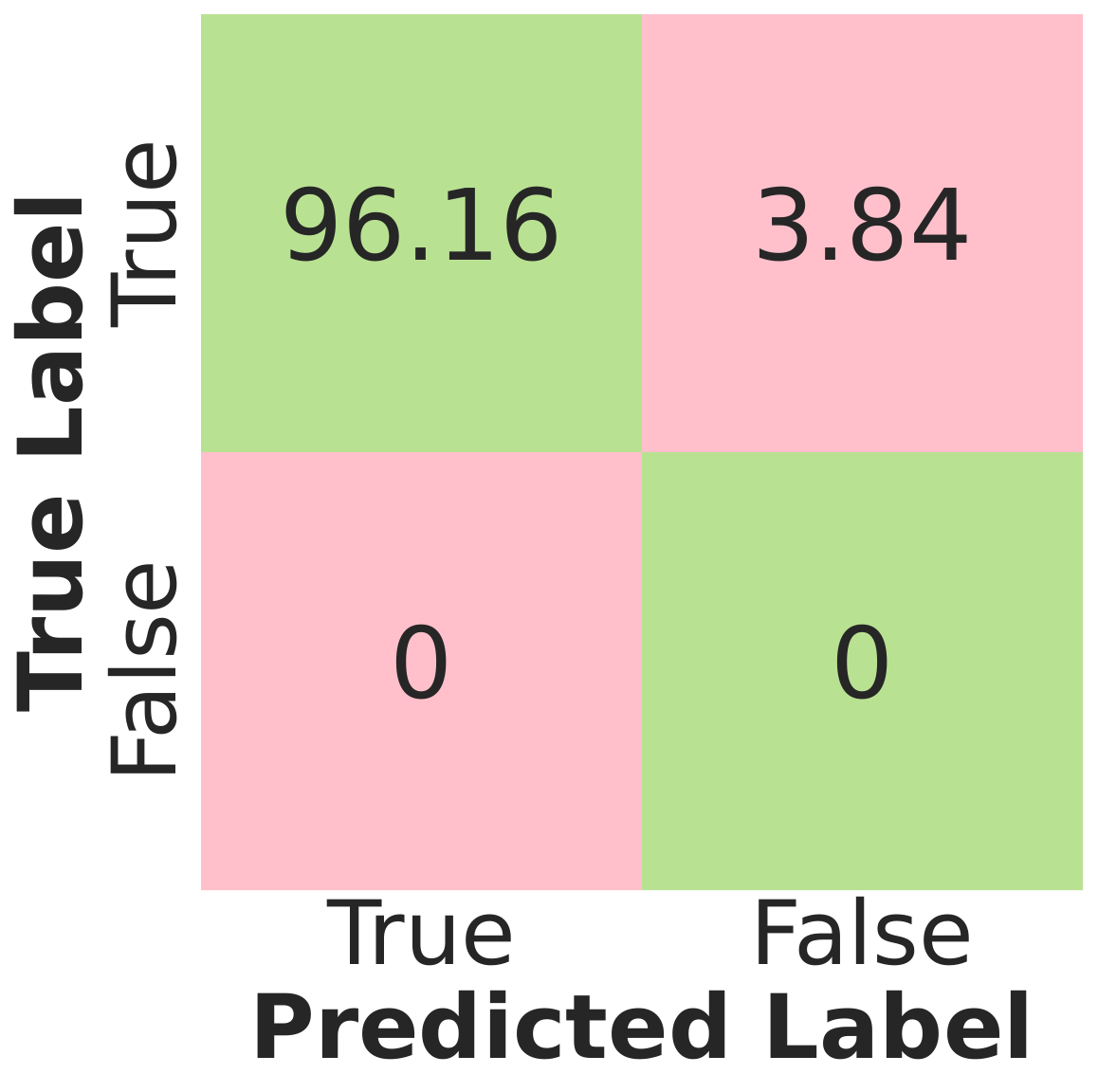}
   \label{fig:yolov8-1-adv}
  }~~
  \subfloat[Medium]{
    \includegraphics[width=0.28\columnwidth]{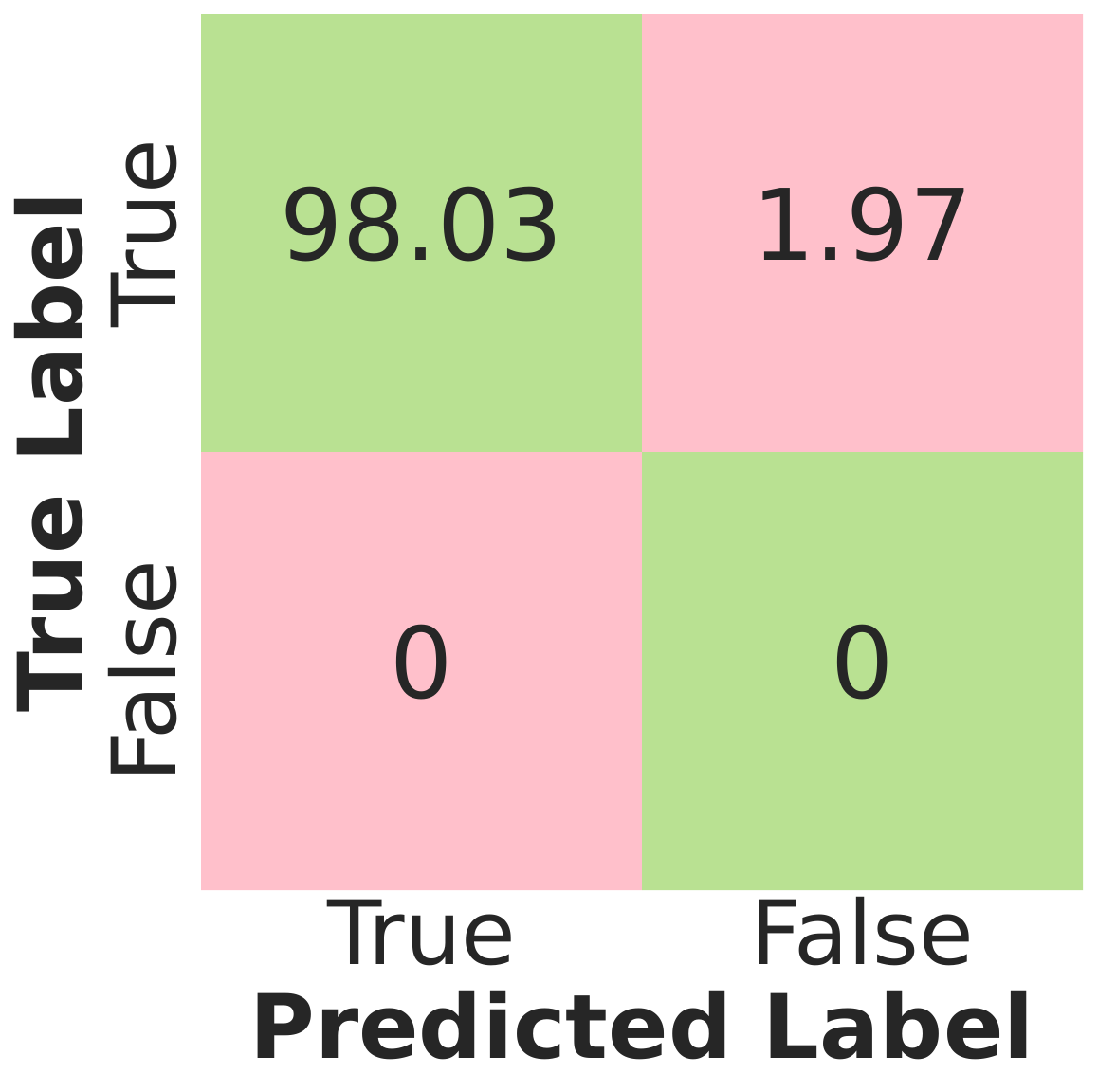}
   \label{fig:yolov8-2-adv}
  }~~
  \subfloat[X-Large]{
   \includegraphics[width=0.28\columnwidth]{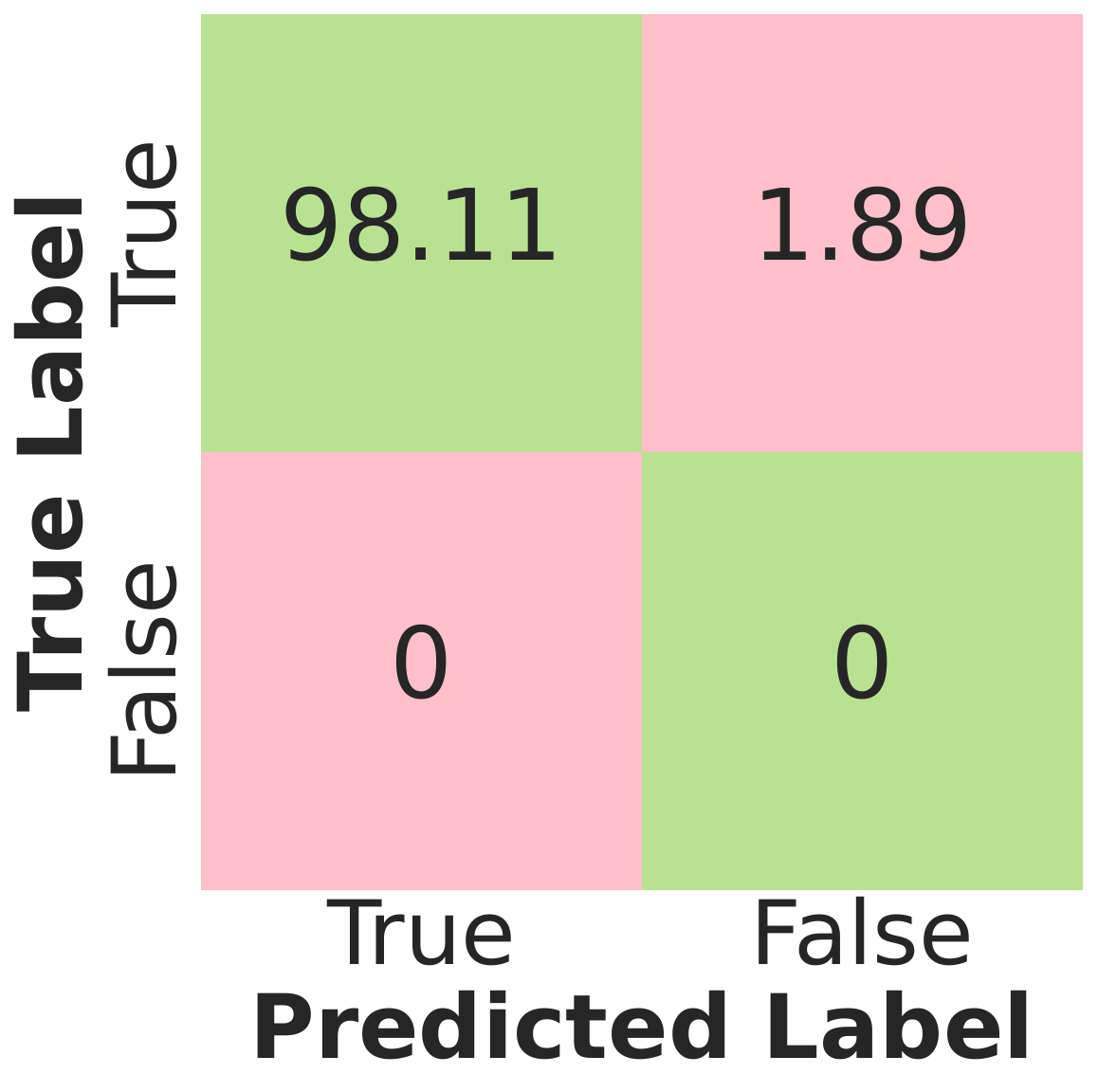}
    \label{fig:yolov8-3-adv}
  }\\
  \subfloat[Nano]{
    \includegraphics[width=0.28\columnwidth]{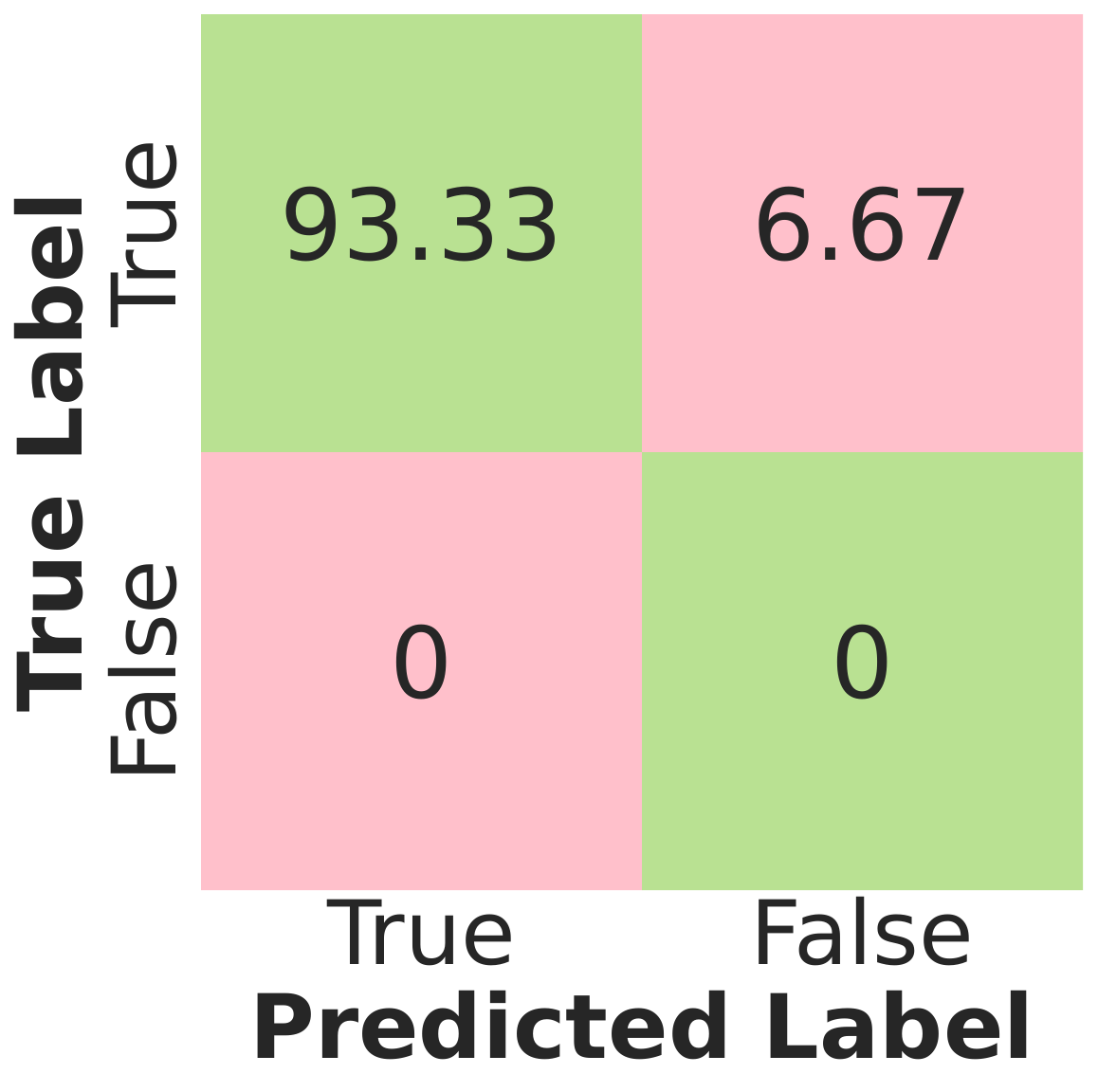}
   \label{fig:yolov11-1-adv}
  }~~
  \subfloat[Medium]{
    \includegraphics[width=0.28\columnwidth]{Scratch-Figs/adversarial/custom_yolov11_m_100_epochs.pdf}
   \label{fig:yolov11-2-adv}
  }~~
  \subfloat[X-Large]{
   \includegraphics[width=0.28\columnwidth]{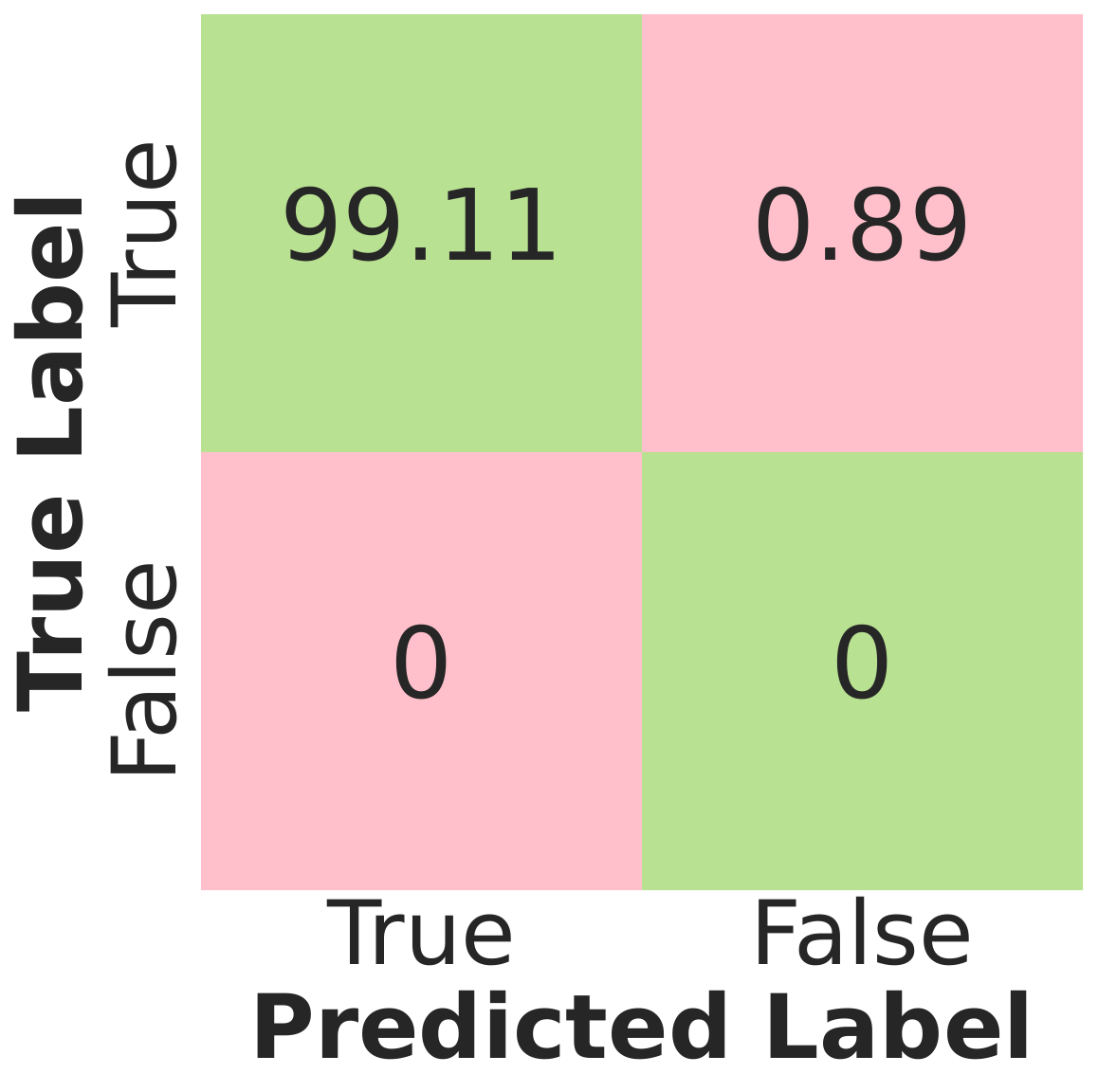}
    \label{fig:yolov11-3-adv}
  }
\caption{Accuracy (in \%) of VIP detection using different sizes of Re-trained (RT) YOLOv8 (top) and YOLOv11 (bottom) on adversarial datasets}
\label{fig:adversarial-accuracy}
\end{figure}



We extensively evaluate the accuracy of Re-trained (RT) YOLO models on a diverse dataset of $23,543$ images and an adversarial dataset of $3,805$ images. To benchmark inference times for all models across devices, we run a subset of approximately $1,000$ images. 
As BodyPose and Monodepth2 models are sourced from existing repositories, we do not report their accuracies. Finally, we present our benchmark study analysis. 
For our results, since there are no false positives, precision equals accuracy.

\begin{figure}[!t]
    \centering
    \subfloat[Inf. time for YOLOv8]{
    \includegraphics[width=0.45\columnwidth]{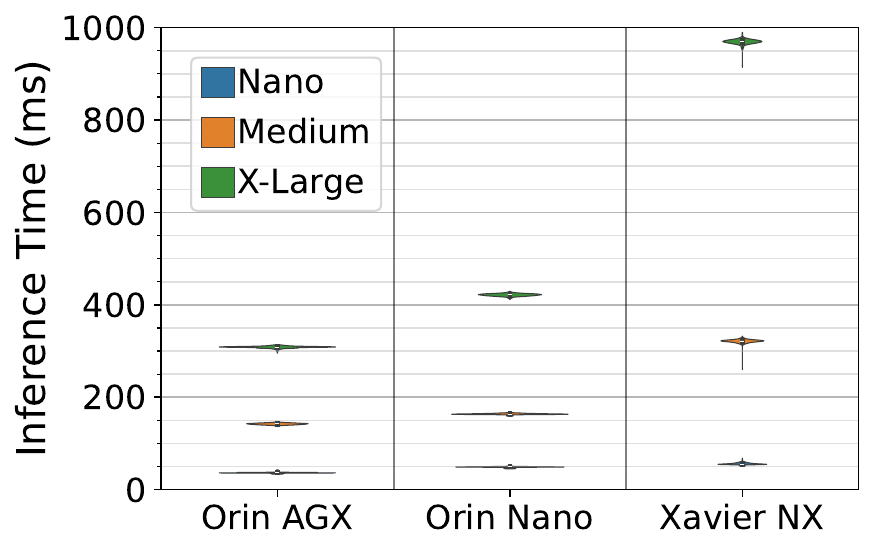}
   \label{fig:inf-time-v8}
  }\hfill
    \subfloat[Inf. time for YOLOv11  ]{
    \includegraphics[width=0.45\columnwidth]{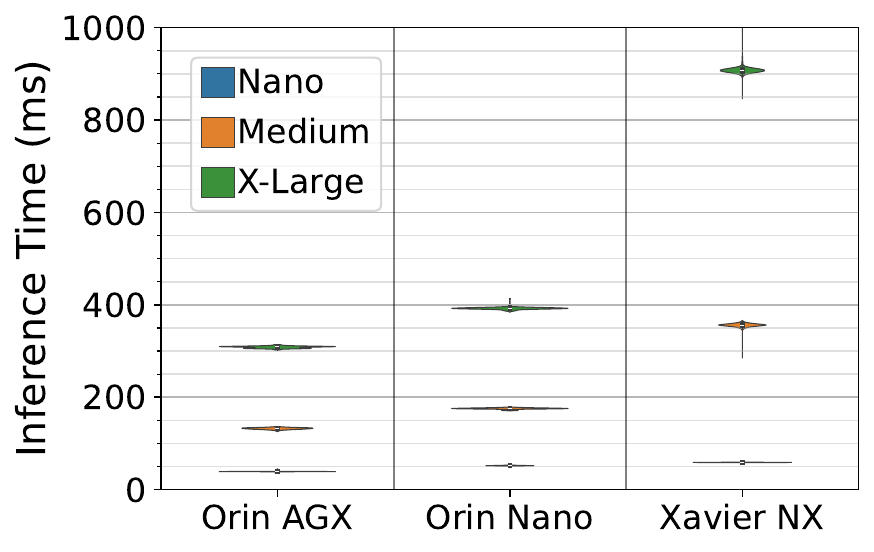}
   \label{fig:inf-time-v11}
  }\\
    \subfloat[Inf. time for Bodypose]{
    \includegraphics[width=0.45\columnwidth]{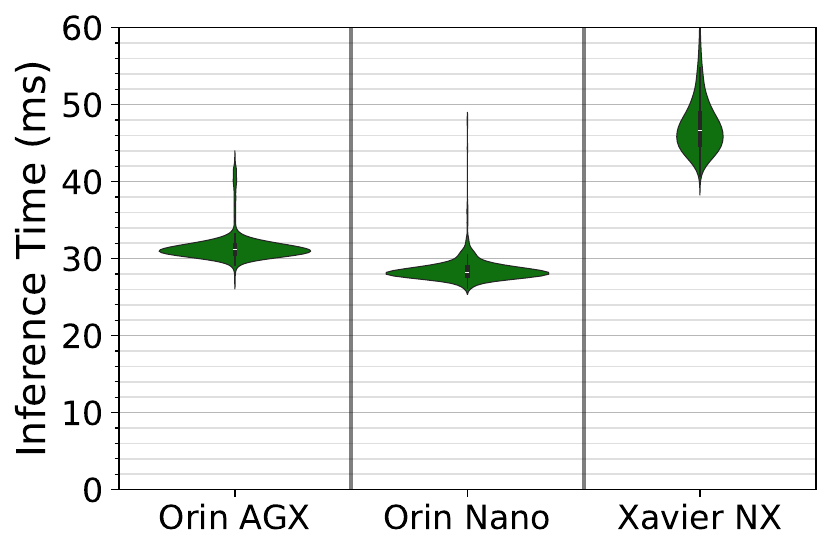}
   \label{fig:inf-bodypose}
  }~~
    \subfloat[Inf. time for Monodepth2]{
    \includegraphics[width=0.45\columnwidth]{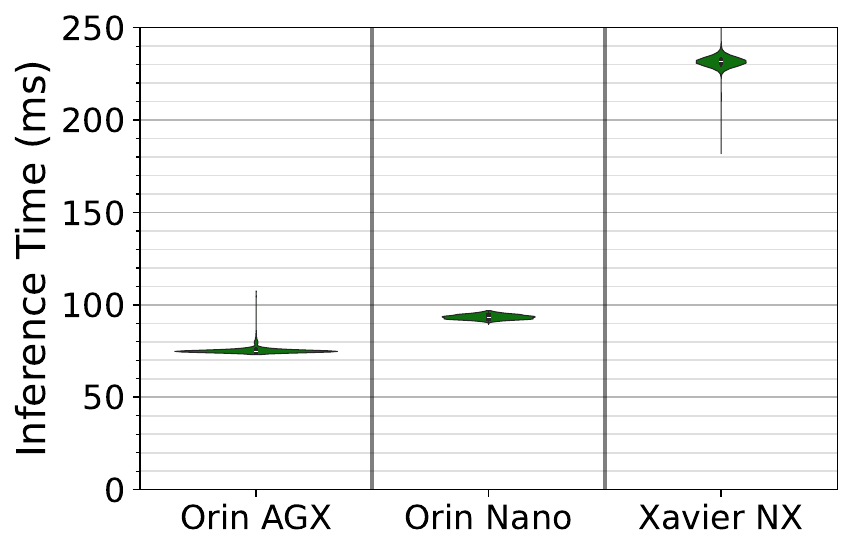}
   \label{fig:inf-monodepth}
  }
\caption{Inference Times on Jetson Edge Accelerators}
\label{fig:inf-edge}
\end{figure}  

\subsubsection{\hilite{Accuracy of YOLOv11 increases marginally compared to YOLOv8 for diverse dataset as the model size increases}} 
As shown in Fig.~\ref{fig:diverse-accuracy}, both re-trained models achieve an accuracy of $\geq98.6\%$, significantly outperforming existing work. Specifically, RT YOLOv8 attains $\approx99\%$ accuracy on diverse datasets. Notably, increasing model size does not yield a significant accuracy improvement. However, RT YOLOv11 achieves $99.49\%$ accuracy for the medium size and $99.27\%$ for the X-large size, demonstrating a marginal advantage over YOLOv8 at comparable sizes. The absence of false positives in our models demonstrates their high precision and robustness in correctly identifying the target object (neon hazard vest) without misclassification. This ensures reliability in real-world scenarios, reducing the risk of incorrect detections that could lead to navigation errors for VIPs.

\subsubsection{\hilite{Accuracy of YOLO models increase with their sizes on the adversarial dataset}}
Figure~\ref{fig:adversarial-accuracy} illustrates the trend of increasing accuracy with model size when tested on adversarial datasets. As observed, the nano model has the lowest accuracy, which improves significantly for the medium size and reaches its peak at the x-large size, $99.11\%$ for YOLOv11 and $98.11\%$ for YOLOv8. This aligns with YOLO’s claim that larger-size models achieve higher accuracy. 

The trend of increasing accuracy with model size is not as evident in the diverse dataset as the diverse dataset provides clear visual cues, allowing even smaller models to achieve high accuracy without needing larger model capacity. In contrast, adversarial datasets present challenging conditions where larger YOLO models leverage increased complexity to enhance robustness. High accuracy on adversarial datasets is particularly valuable, as most of the models often fail in such scenarios, making robustness a key measure of real-world effectiveness.

\begin{figure}[!t]
    \centering
    \includegraphics[width=0.75\columnwidth]{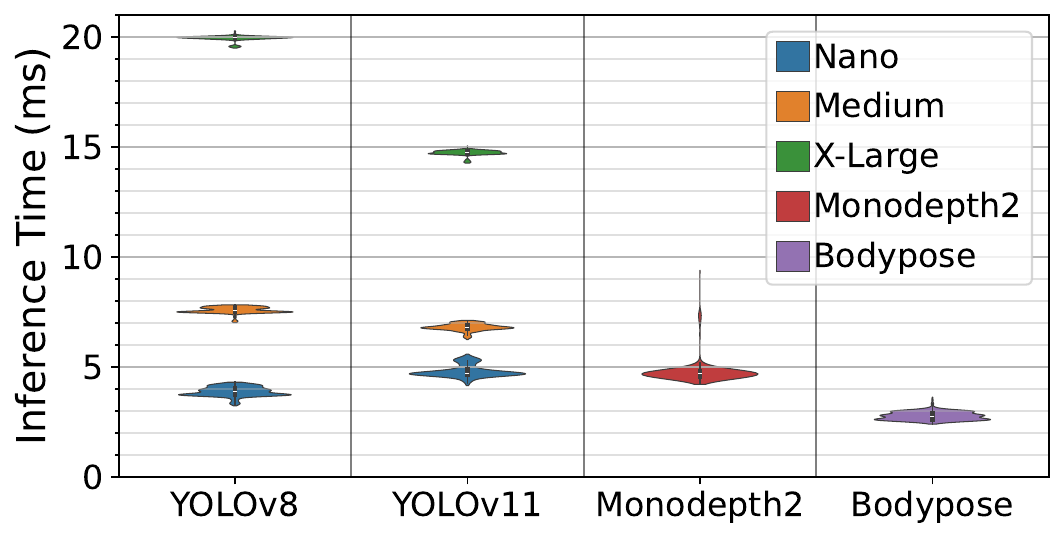}
\caption{Inference Times on RTX 4090 GPU workstation}
\label{fig:inf-cloud}
\end{figure} 

\subsubsection{\hilite{Inference time for models on edge depends on model size and device specifications}}

Figure~\ref{fig:inf-edge} presents the inference time per frame for various YOLO model sizes, along with Bodypose and Monodepth2 models, on edge devices. As detailed in Table~\ref{tab:computespecs}, Orin AGX (o-agx) is the most powerful device with $2048$ CUDA cores, followed by Orin Nano (o-nano) with $1024$ cores, and Xavier NX (nx) with only $384$ cores. Given that the Ampere architecture is more efficient and scalable than Volta, we observe the fastest inference on o-agx, followed by o-nano, with nx being the slowest. For YOLO models, both nano and medium variants achieve inference times of $\leq200$~ms, while x-large models remain under $500$~ms. However, on nx, only the nano model stays within $200$~ms, whereas x-large models exhibit significantly higher inference times, reaching up to $989$ ms.

We observe a similar trend in Fig.~\ref{fig:inf-bodypose} and Fig.~\ref{fig:inf-monodepth}. Bodypose model has a median inference time ranging between $28 - 47$~ms on these devices, whereas Monodepth2 has a higher inference time of $75 - 232$~ms. These can be tied back to the model sizes and number of parameters in Table~\ref{tab:dnn-model-specs}.

\subsubsection{\hilite{Inference time for all models are $\leq25$~ms on GPU workstation}}
With approximately $8\times$ more CUDA cores than Orin AGX, the RTX 4090 demonstrates a substantial improvement in inference times across all models, shown in Fig.~\ref{fig:inf-cloud}. The nano and medium sizes of both YOLO models, along with Bodypose and Monodepth2, achieve inference times within $10$~ms per frame, while the x-large models remain under $20$~ms—approximately $50\times$ faster than on Xavier NX. This highlights the advantage of leveraging GPU cloud resources alongside resource-constrained edge devices for better collaboration in real-time applications, where larger models with higher accuracy can be hosted on the workstation, and smaller models with lower accuracy can be hosted on edge devices. Overall, we observe that all models achieve an inference time of $\leq25$~ms per frame on the workstation.


\section{Conclusions and Future Work}\label{sec:conclusions}
In this work, we proposed Ocularone-Bench, a benchmark suite designed for real-time VIP navigation assistance. Our benchmarks include a curated dataset of individuals wearing hazard vests in diverse and adversarial environments, retrained YOLO models achieving up to $99.49\%$ accuracy, and comprehensive inference time benchmarks across various edge accelerators and high-end GPU workstations.

Future work includes expanding the dataset with more diverse real-world scenarios, integrating multi-modal sensing (LiDAR, thermal imaging), and developing accuracy-aware adaptive deployment strategies for seamless execution across edge-cloud environments.

\section*{Acknowledgements}
The authors would like to thank members of Dream:Lab, including Ansh Bhatia, Swapnil Padhi, Prince Modi and Akash Sharma for their assistance with the paper. 

\bibliographystyle{IEEEtran}
\bibliography{ipdps}

\begin{thebibliography}{10}
\providecommand{\url}[1]{#1}
\csname url@rmstyle\endcsname
\providecommand{\newblock}{\relax}
\providecommand{\bibinfo}[2]{#2}
\providecommand\BIBentrySTDinterwordspacing{\spaceskip=0pt\relax}
\providecommand\BIBentryALTinterwordstretchfactor{4}
\providecommand\BIBentryALTinterwordspacing{\spaceskip=\fontdimen2\font plus
\BIBentryALTinterwordstretchfactor\fontdimen3\font minus \fontdimen4\font\relax}
\providecommand\BIBforeignlanguage[2]{{%
\expandafter\ifx\csname l@#1\endcsname\relax
\typeout{** WARNING: IEEEtran.bst: No hyphenation pattern has been}%
\typeout{** loaded for the language `#1'. Using the pattern for}%
\typeout{** the default language instead.}%
\else
\language=\csname l@#1\endcsname
\fi
#2}}

\bibitem{whoStats}
{World Health Organization}, ``\href{https://www.who.int/news-room/fact-sheets/detail/blindness-and-visual-impairment}{Blindness and visual impairment fact sheets},'' August 2023.

\bibitem{wewalk}
\BIBentryALTinterwordspacing
WeWALK, ``Wewalk smart cane,'' 2025. [Online]. Available: \url{https://wewalk.io/en/product/}
\BIBentrySTDinterwordspacing

\bibitem{suman2023chi}
S.~Raj, S.~Padhi, and Y.~Simmhan, ``Ocularone: Exploring drones-based assistive technologies for the visually impaired,'' in \emph{Extended Abstracts of the CHI Conference on Human Factors in Computing Systems}, 2023.

\bibitem{roboflow_vest_dataset}
\BIBentryALTinterwordspacing
{RoboFlow Universe}, ``Vest detection datasets,'' Online, 2025. [Online]. Available: \url{https://universe.roboflow.com/search?q=class%3Avest+model+object+detection}
\BIBentrySTDinterwordspacing

\bibitem{ahmad2024sh17}
H.~M. Ahmad and A.~Rahimi, ``Sh17: A dataset for human safety and personal protective equipment detection in manufacturing industry,'' \emph{Journal of Safety Science and Resilience}, 2024.

\bibitem{hazard-vest_dataset}
\BIBentryALTinterwordspacing
Tello, ``hazard-vest dataset,'' aug 2023. [Online]. Available: \url{https://universe.roboflow.com/tello-8ckdt/hazard-vest}
\BIBentrySTDinterwordspacing

\bibitem{10.1007/978-981-96-0805-8_11}
D.~K. Alqahtani, M.~A. Cheema, and A.~N. Toosi, ``Benchmarking deep learning models for object detection on edge computing devices,'' in \emph{Service-Oriented Computing}.\hskip 1em plus 0.5em minus 0.4em\relax Springer Nature Singapore, 2025.

\bibitem{raj2024adaptiveheuristicsschedulingdnn}
\BIBentryALTinterwordspacing
S.~Raj, R.~Mittal, H.~Gupta, and Y.~Simmhan, ``Adaptive heuristics for scheduling dnn inferencing on edge and cloud for personalized uav fleets,'' 2024. [Online]. Available: \url{https://arxiv.org/abs/2412.20860}
\BIBentrySTDinterwordspacing

\bibitem{ultralytics2025}
\BIBentryALTinterwordspacing
Ultralytics, ``Ultralytics yolo models documentation,'' 2025. [Online]. Available: \url{https://docs.ultralytics.com/models/}
\BIBentrySTDinterwordspacing

\bibitem{nvidia_trt_pose}
N.~AI-IOT, ``trt\_pose: Real-time pose estimation accelerated with tensorrt,'' \url{https://github.com/NVIDIA-AI-IOT/trt_pose}, 2023.

\bibitem{monodepth2github}
I.~Niantic, ``\href{https://github.com/nianticlabs/monodepth2}{Monodepth2: Monocular depth estimation from a single image},'' 2019.

\end{thebibliography}

\end{document}